\documentclass[a4paper]{jpconf}
\usepackage{iopams}
\usepackage{cite}
\usepackage[dvips]{graphicx}
\usepackage{pstricks}

\usepackage{epsfig}

\global\arraycolsep=2pt
\newcommand{\ben}{\begin{equation*}}
\newcommand{\een}{\end{equation*}}
\newcommand{\bean}{\begin{eqnarray*}}
\newcommand{\eean}{\end{eqnarray*}}
\newcommand{\nn}{\nonumber}
\newcommand{\be}{\begin{equation}}
\newcommand{\ee}{\end{equation}}
\newcommand{\bea}{\begin{eqnarray}}
\newcommand{\eea}{\end{eqnarray}}

\begin{document}
\title{Recent Developments in the Casimir Effect}
\author{K A Milton\footnote{The work reported here was
done in  collaboration with  I Brevik, S {\AA}dn{\o}y Ellingsen and J S H\o ye
(\sref{sec1});  S Fulling and A Romeo (\sref{sec2});
I Cavero-Pel\'aez, P Parashar, K V Shajesh and 
J Wagner (sections \ref{sec2} and \ref{sec3}).}}

\address{
Oklahoma Center for High Energy Physics, and
H. L. Dodge Department of Physics and Astronomy,
University of Oklahoma, Norman, OK 73019 USA}

\ead{milton@nhn.ou.edu}

\begin{abstract}
In this talk I  review various developments in the past year
concerning quantum vacuum energy, the Casimir effect.  In particular,
there has been continuing controversy surrounding the temperature
correction to the Lifshitz formula for the Casimir force between
real materials, be they metals or semiconductors. Consensus has emerged
as to how Casimir energy accelerates in a weak gravitational field;
quantum vacuum energy, including the divergent parts which renormalize
the masses of the Casimir plates, accelerates indeed according to the
equivalence principle.  Significant development has been forthcoming
in applying the multiple scattering formalism to describe the interaction
between nontrivial objects.  In weak coupling, closed-form expressions
for the Casimir force between the bodies,
which for example reveal significant discrepancies from the naive
proximity force approximation, can be achieved in many cases.
\end{abstract}

\section{Introduction}
There has been burgeoning interest in the Casimir effect \cite{casimir48}, 
the quantum
force between macroscopic bodies due to zero-point fluctuations, as
witnessed by the multitude of international conferences and workshops
devoted to the subject, including two in Brazil back-to-back.  This
interest reflects a great deal of progress in the subject both from
the experimental and theoretical sides.  I will talk about three topics
my collaborators and I have been involved with during the past several
months, which mirror these developments: the controversy concerning the
temperature dependence of real materials, such as metals and semiconductors;
the question of how the Casimir energy associated with
parallel Casimir plates is accelerated by gravity;
and the achievement of exact closed-form results for the interaction 
of bodies subject to weak coupling.

\section{Temperature Controversy}
\label{sec1}
\subsection{Introduction}
There is a continuing controversy surrounding 
the question of how to incorporate thermal
corrections into the Casimir force between real metal
plates \cite{brevik06,klimchitskaya06}.  The procedure to do
this seems straightforward. The thermal Green's functions
must be periodic in imaginary time, with period $\beta=1/T$.
This implies a Fourier series decomposition, rather
than a Fourier transform, where 
we have a sum over Matsubara frequencies $\zeta_m=2\pi m T$.

Thus, the Casimir (Lifshitz) \cite{lifshitz55} pressure
between parallel semi-infinite nonmetallic plates
separated by a distance $a$ is given by
\be
P^T=-\frac{T}\pi\sum_{i={\rm TE,TM}}
\sum_{m=0}^\infty {}'\int_{\zeta_m}^\infty \rmd\kappa
\,\kappa^2\frac1{(r^{i})^{-2}\rme^{2\kappa a}-1},
\ee
where the prime means that the $m=0$ term is counted with half weight,
in terms of the reflection coefficients
\numparts
\bea
r^{\rm TE}=\frac{-\sqrt{1+\frac{\zeta^2}{\kappa^2}(\varepsilon(\rmi\zeta)
-1)}+1}
{\sqrt{1+\frac{\zeta^2}{\kappa^2}(\varepsilon(\rmi\zeta)-1)}+1},\\
r^{\rm TM}=\frac{-\sqrt{1+\frac{\zeta^2}{\kappa^2}(\varepsilon(\rmi\zeta)-1)}+
\varepsilon(\rmi\zeta)}
{\sqrt{1+\frac{\zeta^2}{\kappa^2}(\varepsilon(\rmi\zeta)-1)}
+\varepsilon(\rmi\zeta)}.
\eea
\endnumparts

Let us rewrite the Lifshitz formula at finite temperature in the form
\be
P^T=\sum_{m=0}^\infty{}'f_m=\int_0^\infty \rmd m\,f(m)-\sum_{k=0}^\infty
\frac{B_{2k}}{(2k)!}f^{(2k-1)}(0),
\ee
where the second equality uses the Euler-Maclaurin sum formula, in terms
of
\be
f(m)=-\frac1{2\pi\beta}\sum_i\int_0^\infty \rmd k_\perp^2\,
\kappa(\zeta_m)(d^i_m)^{-1},
\label{fofm}
\ee
where 
\be
d_m^i=(r^i)^{-2} \rme^{2\kappa a}-1,
\ee and $\kappa^2=k_\perp^2+\zeta^2$.

Evidently, for the Drude model,
where the permittivity has the form
\be\varepsilon(\rmi \zeta)=1+\frac{\omega_p^2}{\zeta(\zeta+\gamma)},
\label{drude}\ee
 or more generally, whenever
\be
\lim_{\zeta\to0}\zeta^2[\varepsilon(\rmi \zeta)-1]=0,
\ee
the TE zero mode vanishes,
\be f^{({\rm TE})}(0)=0.
\ee
This leads to a physical discontinuity in the TE mode, between the
$m=0$ and $m=1$ Matsubara frequencies.
Naively, this implies an additional linear term in the pressure
at low temperatures:
\be
P^T\sim P^{T=0}+\frac{\zeta(3)}{8\pi a^3}T, \qquad aT\ll1.\label{linear}
\ee
Exclusion of the TE zero mode will also reduce the linear temperature
dependence expected at high temperatures,
\be
P^T\sim -\frac{\zeta(3)}{8\pi a^3}T,\qquad aT\gg1,\label{hightlinear}
\ee
one-half the usual ideal metal result.  This anomaly was first pointed
out by Bostr\"om and Sernelius \cite{bostrom}.  See also
\cite{Hoye:2002at}.

Several other recent papers also lend support to our point of view. 
 Jancovici and \v Samaj \cite{jancovici04,jancovici} and  Buenzli and Martin
\cite{buenzli05,buenzli}
have examined the Casimir force between ideal-conductor walls with
emphasis on the high-temperature limit.  Not surprisingly, ideal inert
boundary conditions are shown to be inadequate, and fluctuations within
the walls, modeled by the classical Debye-H\"uckel theory, determine the
high temperature behavior.
 The linear in temperature  behavior
of the Casimir force is found to be reduced  by a factor of two
from the behavior predicted by an ideal metal, just as seen above in  
\eref{hightlinear}.   This is precisely the
signal of the omission of the $m=0$ TE mode.
Further support for our conclusions can be found in the work
of Sernelius \cite{sernelius}, who calculated the van der Waals-Casimir
force between gold plates using the Lindhard or random phase approximation
dielectric function.
The central theme of his work is to describe the thermal Casimir
effect in terms of spatial dispersion. 
He also finds that for large separations the force is
one-half that of the ideal metal. 
Sernelius shows
that, for arbitrary separation between the plates, the spatial-dispersion
results nearly exactly coincide with the local dissipation-based results.

\begin{figure}
  \begin{center}
    \includegraphics[width=4.2in]{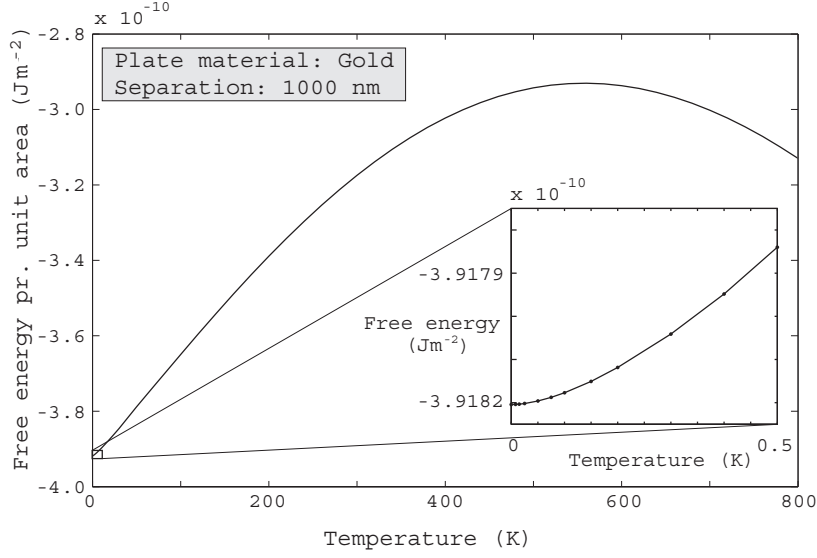}
    \caption{Numerical evaluation of the  free energy between two gold
    halfspaces as a function of temperature. The inset gives details for low
    $T$.}
    \label{fig_F(T)}
  \end{center}
\end{figure}

Thus, it is very hard to
see how the corresponding modification of the low-temperature behavior
can be avoided.  However, because a linear $T$ dependence in the
pressure implies a linear $T$ dependence in the free energy,
this seems to imply a violation of the
Nernst heat theorem, the third law of thermodynamics, because then
a naive calculation implies $S(T=0)\ne0$.
However, we \cite{Brevik:2007wj} and others have shown that for real metals,
where $\gamma$ has a residual value at zero temperature,
the free energy per area has a vanishing slope at the origin. Indeed, in the
Drude model, 
the free energy per area has the behavior \cite{hoye07,brevik04}
\be
F=F_0+T^2 \frac{\omega_p^2}{48\gamma}(2\ln2-1),
\ee
for sufficiently low temperatures. The $T^{5/2}$ and $T^3$ corrections to this
were calculated last year in \cite{hoye07,Brevik:2007wj}
This is illustrated in \fref{fig_F(T)}, taken from \cite{Brevik:2007wj},
which shows the numerical evaluation of the free energy for gold.

\subsection{Experimental constraints}
 The difficulty is that,
experimentally, it is not easy to perform Casimir force measurements at other
than room temperature, so 
current constraints on the theory all come from room
temperature experiments.  Then all one can do is compare the theory at
room temperature with the experimental results, which must be corrected
for a variety of effects, such as surface roughness, finite conductivity,
and patch potentials.
{\em There are no direct experiments which observe $T$ dependence
in metals!} (The temperature dependence of the Casimir-Polder force
between a Bose-Einstein condensate and a dielectric substrate was
measured in 2006 \cite{cornell}.)

Although the earlier experiment by Lamoreaux \cite{lamor} was carried out at 
such large distances ($\sim1\,\mu$m) where the temperature correction proposed
here would be large ($\sim15$\%) \cite{Brevik:2005uh}.
there are reasons to think that the
uncertainties due to such things as surface roughness and patch potentials
were so large as to afford only limited accuracy in his experiment.
There has been a great number of high quality experiments during
the last decade, revealing a number of facets of the quantum vacuum
energy \cite{mohideen98,roy99,harris00,ederth00,chan01,chen02,
bressi02,iannuzzi04,lisanti05, chen07}.
As for the temperature effect, the most
stringent test is given by the precision experiments of
Decca \etal \cite{decca05, bezerra06}, where the agreement of the data
with the zero-temperature theory seems to rule out a large thermal
correction such as Bostr\"om and Sernelius \cite{bostrom} and we have proposed.
It seems to us that there may be a number of effects that have not
been properly accounted for before one can conclude that no large
temperature effect is present, such as the shielding effect recently
proposed by Pitaevskii \cite{pitaevskii08}.
\subsection{Anomaly for semiconductors}

Recently Klimchitskaya, Geyer and Mostepanenko
\cite{geyer,klim,geyer2,klim2} have observed there is a similar
anomaly, now affecting the TM reflection coefficient, for dielectrics
and semiconductors.
Here is a simple way to understand that argument.
Suppose we model a dielectric with some small conductivity
by the permittivity function
\be
\varepsilon(i\zeta)=1+\frac{\bar\varepsilon-1}{1+\zeta^2/\omega_0^2}+\frac{4\pi
\sigma}\zeta.
\ee
(The only essential point is that as $\zeta\to0$, $\varepsilon\to
\bar\varepsilon$
if $\sigma=0$, otherwise $\varepsilon\to\infty$.)

For a dielectric the TE reflection coefficient is continuous,
 but if there is
vanishingly small (but not zero) conductivity the TM coefficient is
discontinuous:
\be
r^{\rm TM}(\zeta=0)=1,\quad \lim_{\zeta\to0}r^{\rm TM}=\frac{\bar\varepsilon-1}
{\bar\varepsilon+1}.
\ee
 This implies a linear term in the pressure
\bea
 P_T^{\rm TM}&=&\frac{T}{2\pi}\int_0^\infty \rmd\kappa\,\kappa^2\left[
\frac1{\left(\frac{\bar\varepsilon+1}{\bar\varepsilon-1}\right)^2 
\rme^{2\kappa a}-1}
-\frac1{\rme^{2\kappa a}-1}\right]\nn\\
&=&\frac{T}{2\pi}\sum_{n=1}^\infty \left[\left(\frac{\bar\varepsilon-1}
{\bar\varepsilon+1}\right)^{2n}-1\right]\int_0^\infty \rmd\kappa \,\kappa^2 
\rme^{-2n\kappa a}
=\frac{T}{8\pi a^3}\left[{\rm Li}_3\left(\left[\frac{\bar\varepsilon-1}
{\bar\varepsilon+1}\right]^2\right)-\zeta(3)\right],
\eea
where the polylogarithmic  function is
\be
{\rm Li}_n(\xi)=\sum_{k=1}^\infty \frac{\xi^k}{k^n}.
\ee
Note that the linear term vanishes for $\bar\varepsilon\to\infty$.  The free
energy term $F$ is obtained from this by multiplying by $a/2$.  Thus at zero
temperature, the entropy is nonzero,
\be
S(0)=-\left(\frac{\partial F}{\partial T}\right)_V=-\frac1{16\pi a^2}
\left[\mbox{Li}_3\left(\left[\frac{\bar\varepsilon-1}{\bar\varepsilon+1}
\right]^2
\right)-\zeta(3)\right].
\ee
Moreover, besides this thermodynamical contradiction, recent experiments
by Mohideen's group \cite{chen07,kmm07,castillo, chen07a} 
show that apparently the conductivity effect must
be excluded for semiconductors with very low carrier concentrations,
but included for semiconductors with high carrier concentrations.
This conundrum has not yet been resolved, although there has been
some progress in understanding what happens when the conductivity remains
small but nonzero at zero temperature \cite{ellingsen08}
---see also Simen \AA dn\o y Ellingsen's talk
in this meeting \cite{ellingsen08a}. We should also mention our joint work on
temperature effects and anomalies \cite{Brevik:2008ry} which, 
while not addressing
the semiconductor anomaly, attempts to survey the underpinnings of
thermal effects in the Casimir effect.

\section{How Does Casimir Energy Fall?}
\label{sec2}
In a series of papers last year \cite{Fulling:2007xa,Milton:2007ar,
Milton:2007hd} we
showed that the total Casimir energy of an apparatus
consisting of parallel plates, including the {\it divergent
parts} of the energy, which renormalize the masses of the plates,
possesses the gravitational mass demanded by the equivalence principle.

This result puts the lie to the naive presumption that zero-point energy is not
 observable.
On the other hand, because of the  severe divergence structure
of the theory, controversy has surrounded it from the beginning.
Sharp boundaries give rise to divergences in the local energy density near
the surface, which may make it impossible to extract meaningful self-energies
of single objects, such as the perfectly conducting sphere considered by
Boyer \cite{Boyer:1968uf}.
These objections have recently been most forcefully presented by
Graham, Jaffe, \etal \cite{Graham:2003ib}, but they date back to
 Deutsch and Candelas \cite{Deutsch:1978sc}.
In fact,
it now appears that these surface divergences can be dealt with successfully
in a process of renormalization,
and that finite self-energies in the sense
of Boyer may be extracted \cite{CaveroPelaez:2006kq,CaveroPelaez:2006rt}.

Gravity couples to the local energy-momentum tensor, and such
surface divergences promise serious difficulties.
We first ask how does the
completely finite Casimir interaction
energy of a pair of parallel conducting plates, as well as the divergent
self-energies of non-ideal plates,
couple to gravity?  Disparate answers had been given in the past
\cite{calloni,bimonte}; thus,
the question, and its answer, turn out to be surprisingly less straightforward
than the reader might suspect!

We will not here address what gravitational field results from the
(in general divergent) Casimir energy.
For a beginning of the renormalization of Einstein's equations
resulting from singular Casimir surface energy densities see 
\cite{Estrada:2008zz}.

Brown and Maclay \cite{brown} showed that, for parallel perfectly
conducting plates separated by a distance $a$ in the $z$-direction,
the electromagnetic stress
tensor acquires the vacuum expectation value between the plates
\be
\langle T^{\mu\nu}\rangle
=\frac{\mathcal{E}_c}a \mbox{diag} (1, -1, -1, 3),
\qquad \mathcal{E}_c=-\frac{\pi^2}{720 a^3}\hbar c.\label{t}\ee
Outside the plates the value of $\langle T^{\mu\nu}\rangle=0$.

Because there are some subtleties here, let us review the 
derivation of \eref{t} for the
case of a conformally coupled scalar (the electromagnetic case differs
by a factor of two).  Actually, the result between the plates,
$0<z<a$, is given
in great detail in \cite{casbook} ($\xi$ is the conformal parameter) 
\bea
\langle T^{\mu\nu}\rangle=
(u_0+u)\mbox{diag}(1,-1,-1,3)
+(1-6\xi)s(z)\mbox{diag}(1,-1,-1,0),\label{bmt}
\eea
where
\be
u_0=-\frac1{12\pi^2}\int_0^\infty \rmd\kappa\,\kappa^3,\qquad
u=-\frac{\pi^2}{1440 a^4},
\ee

Note that $u_0$ is a divergent constant, independent of $a$,
and is present (as we shall see) both inside and outside the plates,
so it does not contribute to any observable force or energy (the force
on the plates is given by the discontinuity of $\langle T_{zz}\rangle$),
and so may be simply disregarded (as long as we are not concerned with
dark energy). Similarly, in terms of the Hurwitz zeta function,
\be
s(z)=-\frac1{16\pi^2a^4}[\zeta(4,z/a)+\zeta(4,1-z/a)],
\ee
which exhibits the universal surface
divergence near the plates,
\be
s(z)\sim -\frac1{16\pi^2 z^4},\qquad z\to 0+,\label{indiv}
\ee
is also unobservable (if we disregard gravity)
because it does not
contribute to the force on the plates, nor does
it contribute to the total energy, since the integral over $s(z)$ between
the plates is independent of the plate separation.  Of course, the best
way to eliminate that term is to choose the  conformal value $\xi=1/6$.

Since the exterior calculation does not appear to be referred to 
in \cite{casbook},
let us sketch the calculation here:  Consider parallel
Dirichlet plates at $z=0$ and $z=a$.
The reduced Green's function satisfies
\be
\left(-\frac{\rmd^2}{\rmd z^2}+\kappa^2\right)g(z,z')=\delta(z-z'),
\ee
where $\kappa^2=k^2-\omega^2=k^2+\zeta^2$.  The solution for $z, z'<0$ is
\be
g(z,z')=-\frac1\kappa e^{\kappa z_<}\sinh \kappa z_>.
\ee
where $z_{<,>}$ means the lesser or greater of $z$ and $z'$.
It is very straightforward to calculate the one-loop expectation value of
the stress tensor from
\be
\langle T^{\mu\nu}\rangle=\left(\partial^\mu\partial^{\prime\nu}-\frac12
g^{\mu\nu}\partial^\lambda\partial_\lambda'\right)\frac1\rmi
G(x,x')\bigg|_{x'=x}
-\xi(\partial^\mu\partial^\nu-g^{\mu\nu}\partial^2)\frac1\rmi G(x,x),
\ee
which follows from
\be
\langle\phi(x)\phi(x')\rangle=\frac1\rmi G(x,x')\label{vev}.
\ee
After integrating over $\omega=\rmi\zeta$ and $\mathbf{k}$,
we find the result ($z<0$)
\be
\langle T^{\mu\nu}\rangle=u_0\mbox{diag}(1,-1,-1,3)-\frac{(1-6\xi)}
{16\pi^2
|z|^4}\mbox{diag}(1,-1,-1,0).
\ee
This is exactly as expected.  The $u_0$ term is the same as inside the
box, so is just the vacuum value, and the second term is the universal surface
divergence seen in \eref{indiv} (independent of plate separation),
which can be eliminated by setting $\xi=1/6$.

Thus, we conclude that the
physical stress tensor VEV is just that found by Brown and Maclay \cite{brown}:
\be
\langle T^{\mu\nu}\rangle =u\,\mbox{diag}(1,-1,-1,3)\theta(z)\theta(a-z).
\label{gravfree}
\ee
in terms of the usual step function.

\subsection{Variational principle}
\label{sec:vp}
Now we turn to the question of the gravitational interaction of
this Casimir apparatus.
It seems this question can be most simply addressed through
use of the gravitational definition of the energy-momentum tensor,
\be
\delta W_m\equiv -\frac12\int(\rmd x) \sqrt{-g}\,\delta g^{\mu\nu}T_{\mu\nu}
=\frac12\int(\rmd x) \sqrt{-g}\,\delta g_{\mu\nu}T^{\mu\nu}.\label{var}
\ee
For a weak field,
\be
g_{\mu\nu}=\eta_{\mu\nu}+2h_{\mu\nu},
\ee
so if we think of turning
on the gravitational field as the perturbation, we can ignore $\sqrt{-g}$.
The gravitational energy, for a static situation, is therefore given by
($\delta W=-\int \rmd t\,\delta E$)
\be
E_g=-\int (\rmd\mathbf{x}) h_{\mu\nu}T^{\mu\nu}.\label{ge}
\ee
Here we use the gravity-free electromagnetic Casimir stress tensor 
\eref{gravfree},
with $u$ now replaced by $\mathcal{E}_c/a$ for the electromagnetic situation.

\begin{figure}
\centering
\epsfig{figure=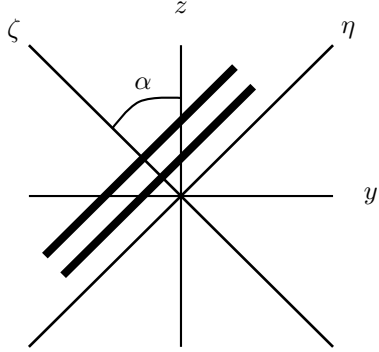}
\caption{Relation between two Cartesian coordinate frames: One attached to
the earth $(x,y,z)$, where $-z$ is the direction of gravity, and one attached
to the parallel-plate Casimir apparatus $(\zeta,\eta,\chi)$, where $\zeta$ is
in the direction normal to the plates.  The parallel plates are
indicated by the heavy lines parallel to the $\eta$ axis.  The $x=\chi$
axis is perpendicular to the page.}
\label{fig1}
\end{figure}

The Fermi metric describes an inertial coordinate system, as we will
discuss further below in \sref{sec3.2}:
\be
h_{00}=-gz,\quad h_{0i}=h_{ij}=0.\label{fm}
\ee
Let us consider a Casimir apparatus of parallel plates
separated by a distance $a$, with transverse dimensions $L\gg a$.
Let the apparatus be oriented at an angle $\alpha$ with respect to the
direction of gravity, as shown in \fref{fig1}.
The Cartesian coordinate
system attached to the earth is $(x, y, z)$, where
$z$ is the direction of $-\mathbf{g}$.  The Cartesian coordinates
associated with the Casimir apparatus 
are $(\zeta,\eta,\chi)$, where $\zeta$
is normal to the plates, and $\eta$ and $\chi$ are parallel to the plates.

The relation between the two sets of coordinates is
\numparts\bea
z&=&\zeta\cos\alpha+\eta\sin\alpha,\\
y&=&\eta\cos\alpha-\zeta\sin\alpha,\\
x&=&\chi.
\eea
\endnumparts
Let the center of the apparatus be located at
\be(\zeta=\zeta_0,\eta=0,\chi=0).\ee

Now we calculate the gravitational energy from \eref{ge},
\be
E_g=\int (\rmd\mathbf{x})gz \langle T^{00}\rangle
=\frac{\mathcal{E}_c}a gL\int_{-L/2}^{L/2}\rmd\eta
\int_{\zeta_0-a/2}^{\zeta_0+a/2}\rmd\zeta(\zeta\cos\alpha+\eta\sin\alpha)\nn\\
=g\mathcal{E}_cL^2\zeta_0\cos\alpha+K,
\ee
where $K$ is a constant, independent of the center of the
apparatus $\zeta_0$, $z_0=\zeta_0\cos\alpha$.
Thus, the gravitational force per area $A=L^2$
on the apparatus is independent of orientation
\be
\mathcal{F}=
\frac{F}A=-\frac{\partial E_g}{A\partial z_0}
=-g\mathcal{E}_c=\epsilon \mathcal{F}_c,
\label{f0}
\ee
a small upward push. Here $\epsilon=\frac13ga$ is a measure of the 
gravitational force relative to the Casimir force $\mathcal{F}_c=-\pi^2/(240
a^4)$.
Note that on the earth's surface,
the dimensionless number $\epsilon$ is very small.  For a plate separation
of $1 \mu$m,
\be
\epsilon=\frac{ga}{3c^2}\sim 10^{-22},
\ee
so the considerations here would appear to be only of theoretical interest.
 The effect is far smaller than the Casimir forces between the plates.

It is a bit simpler to use the energy formula \eref{ge} to calculate the
force by considering the variation in the gravitational energy
directly, as we illustrate by considering a mass point at the origin:
\be
T^{\mu\nu}=m\delta(\mathbf{r})\delta^{\mu0}\delta^{\nu0}.\label{mp}
\ee
If we displace the particle rigidly upward by an amount $\delta z_0$, the
change in the metric is $\delta h_{00}=-g \delta z_0$.
This implies a change in the energy, exactly as expected:
\be
\delta E_g=-m\left(-g \delta z_0\right)=mg\delta z_0.
\ee

Now we repeat this calculation for the Casimir apparatus. The gravitational
force per area on the rigid apparatus is
\be
\mathcal{F}=-\frac{\delta \mathcal{E}_g}{\delta z_0}=
- g\mathcal{E}_c,\label{f1}
\ee
again the same result found in \eref{f0}, which
is 1/4 that found by Bimonte \etal \cite{bimonte}. 
It does, however, agree with one of the results found in the
earlier paper by the same collaboration \cite{calloni},
who now completely agree with our calculation \cite{Bimonte:2008zv}.
Our result is further completely consistent with the
principle of equivalence, and with one result of
Jaekel and Reynaud \cite{jaekel}.

In electrodynamics, one defines the field (the 4-vector potential) by
\be
\delta W=\int (\rmd x) \delta J^\mu A_\mu.\label{emvar}
\ee
From this, one can compute the force on a charge by considering the
displacement of a static charge,
\be
J^\mu=e\delta^{\mu0}\delta(\mathbf{r}),
\ee
that is,
\be
\delta J^\mu=e\delta^{\mu0}\left(-\delta\mathbf{r}\cdot\bnabla\right)\delta
(\mathbf{r}).\ee
Inserting this in the above and integrating by parts we get
\be
\delta W=\int \rmd t\int (\rmd\mathbf{r}) e\left(-\bnabla A^0\right)
\cdot \delta \mathbf{r}\,\delta (\mathbf{r}).\ee
Because the electric field is 
$E^i=F^{0i}=\partial^0 A^i-\partial^i A^0,$
and a total derivative on time is irrelevant, we deduce the force on the
charge to have the expected form:
\be
-\frac{\delta E}{\delta \mathbf{r}}=\mathbf{F}=e\mathbf{E}.
\ee

We therefore should be able to proceed similarly with vacuum energy, starting
from the definition of the gravitational field
\be
\delta W=\int (\rmd x) \delta T^{\mu\nu} h_{\mu\nu}.\label{2ndmethod}
\ee
Again, check this for the force on a mass point, described by \eref{mp},
so
\be
\delta E_g=- \int
(\rmd\mathbf{r}) m \left(-\delta\mathbf{r}\cdot\bnabla\right)
\delta(\mathbf{r})h^{00}=- m\delta\mathbf{r}\cdot \bnabla h^{00}.
\ee
Since $h^{00}=-gz$, we conclude
\be
-\frac{\delta E}{\delta \mathbf{r}}=\mathbf{F}=-mg \hat z.
\ee

For the same constant field the force on a Casimir apparatus
is obtained from the change in the energy density
\be
\langle T^{00}\rangle=\frac{\mathcal{E}_c}a\theta(a/2-\zeta+\zeta_0)
\theta(\zeta-\zeta_0+a/2),
\ee
that is, recalling that $z_0=\zeta_0\cos\alpha$, we obtain the
change of the energy density under a rigid displacement $\delta z_0$:
\be
\delta \langle T^{00}\rangle=\frac{\mathcal{E}_c}a\delta z_0\frac1{\cos\alpha}
\left[\delta (\zeta-\zeta_0-a/2)-\delta (\zeta-\zeta_0+a/2)\right],
\label{varst}
\ee
which yields the gravitational force, since
$h^{00}=-g (\zeta\cos\alpha+\eta\sin\alpha)$,
\be
\mathcal{F}=-\frac{\delta E_g}{A\delta z_0}=\frac{F}A=\frac{\mathcal{E}_c}a
\frac1{\cos\alpha}
 h^{00}\bigg|_{\zeta =\zeta_0-a/2}^{\zeta=\zeta_0+a/2}=
-g\mathcal{E}_c.
\ee

\subsection{Inertial coordinates}
\label{sec3.2}

However, one might think that the above metric, 
while sufficing for massive Newtonian objects,
might seem inappropriate for photons, near the surface
of the earth.  Rather, shouldn't we use the perturbation of
the Schwarzchild metric,  for weak fields ($GM/r\ll1$)
either in its original or isotropic form?
Doing so would give rise to a different force on the Casimir apparatus.
The reason we get different answers in different coordinate systems
reflects the fact that our starting point is not gauge invariant.
Under a coordinate redefinition, which for weak fields is a gauge
transformation of $h_{\mu\nu}$ \cite{schwinger}
\be
h_{\mu\nu}\to h_{\mu\nu}+\partial_\mu\xi_\nu+\partial_\nu\xi_\mu,
\ee
where $\xi_\mu$ is a vector field,  the interaction $W$
is invariant only if the
stress tensor is conserved, $\partial_\mu T^{\mu\nu}=0.$  Otherwise, there
is a change in the action,
\be
\Delta W=-2\int (\rmd x)\xi_\nu \partial_\mu T^{\mu\nu}.
\ee

Now in our case (where we make the finite size of the plate explicit,
but ignore edge effects because $L\gg a$)
\bea
T^{\mu\nu}=\frac{\mathcal{E}_c}a\mbox{diag}(1,-1,-1,3)
\theta
\left(\zeta-\zeta_0+\frac{a}2\right)
\theta\left(\frac{a}2-\zeta+\zeta_0\right)\nn\\
\times\theta(\eta+L/2)\theta(L/2-\eta)\theta(\chi+L/2)\theta(L/2-\chi).
\label{explicitt}
\eea
Thus the nonzero components of $\partial_\mu T^{\mu\nu}$ are
\numparts
\bea
\partial_\mu T^{\mu\zeta}&=&\frac{3\mathcal{E}_c}a
\left[\delta(\zeta-\zeta_0+a/2)-\delta(\zeta-\zeta_0-a/2)\right]\theta\dots,\\
\partial_\mu T^{\mu\eta}&=&-\frac{\mathcal{E}_c}a\left[\delta(\eta+L/2)-\delta
(\eta-L/2)\right]\theta\dots,\\
\partial_\mu T^{\mu\chi}&=&-\frac{\mathcal{E}_c}a\left[\delta(\chi+L/2)-\delta
(\chi-L/2)\right]\theta\dots,
\eea
\endnumparts
where $\theta\dots$ refer to the remaining step functions.
Therefore, the change in the energy obtained from $\Delta W$ is
\bea
\Delta E_g=\frac{6\mathcal{E}_c}{a}\int \rmd\eta \,\rmd\chi \left[\xi_\zeta(
\zeta_0-a/2,\eta,\chi)-\xi_\zeta(\zeta_0+a/2,\eta,\chi)\right]\nn\\
\mbox{}-\frac{2\mathcal{E}_c}a\!
\int \rmd\zeta\, \rmd\chi \left[\xi_\eta(\zeta,
-L/2,\chi)-\xi_\eta(\zeta,L/2,\chi)\right]\nn\\
\mbox{}-\frac{2\mathcal{E}_c}a
\int \rmd\zeta \,\rmd\eta \left[\xi_\chi(\zeta,
\eta, -L/2)-\xi_\chi(\zeta,\eta,L/2)\right].\label{de}
\eea

Since we have demonstrated that the gravitational force on a Casimir
apparatus is not a gauge-invariant concept, we must ask if there is
any way to extract a physically meaningful result. There seem to be
two possible ways to proceed.  Either we add another interaction, say
a fluid exerting a pressure on the plates, resulting in a total stress
tensor that is conserved, or  we find a physical basis for believing that
one coordinate system is more realistic than another.
The former procedure
is undoubtedly more physical, but will yield model-dependent results.
The latter apparently has a natural solution.

A Fermi coordinate system is the general relativistic generalization of
an inertial coordinate frame.  Such a system has been given in \cite{marzlin}
for a resting observer in the field of a static mass distribution.
For our case, where the gravitational potential is $\phi=gz$ 
(apart from an irrelevant additive constant),
we recover the Fermi coordinate metric \eref{fm} for a gravitating body,
\be
\rmd s^2=-(1+2gz)\rmd t^2+\rmd\mathbf{r}^2.
\ee
This is {\it a priori\/} obvious, because in this coordinate
system, coordinate lengths don't depend on $z$.
The Fermi coordinate system may be immediately derived
from the Schwarzchild coordinates, far from the gravitating body
(so that $GM/r\ll1$), by a simple coordinate redefinition.

\subsection{Alternative derivation}

For an alternative derivation of the force on the Casimir apparatus,
we start from the variational definition of the stress tensor \eref{var},
and consider a general coordinate transformation,
\be
x^{\prime\mu}=x^\mu+\delta x^\mu,
\ee
so that
\be
g'_{\mu\nu}(x')=g'_{\mu\nu}(x)+\partial_\lambda g_{\mu\nu} \delta x^\lambda
= g_{\mu\nu}(x)+\delta g_{\mu\nu}(x),
\ee
where
\be
\delta g_{\mu\nu}=\delta x^\lambda \partial_\lambda g_{\mu\nu}+g_{\alpha \nu}
\frac{\partial \delta x^\alpha}{\partial x^\mu}
+g_{\mu\beta}\frac{\partial \delta x^\beta}{\partial x^\nu}.\label{gct}
\ee

For a rigid translation, $\delta x^\lambda$ is a constant, so only the first
term here is present, which gives the result found above in \sref{sec:vp}.
However, if we don't make this restriction, we obtain
\bea
\delta W=-\int_\Omega
(\rmd x)\left[\partial_\nu\left(\sqrt{-g}T^\nu{}_\lambda\right)
-\frac12\sqrt{-g}T^{\mu\nu}\partial_\lambda g_{\mu\nu}\right]\delta x^\lambda
+\int_{\partial\Omega} \rmd\Sigma_\nu
\sqrt{-g}T^\nu{}_\lambda\delta
x^\lambda,\label{intbyparts}
\eea
where the surface term results from integration by parts.
Again, notice for $\delta x^\lambda$ constant, the first and third terms
identically cancel.
If we take the region $\Omega$ to be all space
 the surface term vanishes (and therefore so
does the first term for a rigid displacement).
A simple calculation indeed shows that the surface term cancels.

\subsection{Rindler coordinates}
Relativistically, uniform acceleration is described by hyperbolic motion
\be
t=\xi\sinh\tau,\quad z=\xi\cosh\tau,
\ee
which corresponds to the metric
\be
\rmd t^2-\rmd z^2-\rmd x_\perp^2=\xi^2\rmd \tau^2-\rmd\xi^2 -\rmd x_\perp^2.
\ee
The d'Alembertian operator has the following form in Rindler coordinates:
\be
-\left(\frac\partial{\partial t}\right)^2
+\left(\frac\partial{\partial z}\right)^2+\nabla_\perp^2
=-\frac1{\xi^2}\left(\frac\partial{\partial\tau}\right)^2+\frac1\xi
\frac\partial{\partial\xi}\left(\xi\frac\partial{\partial\xi}\right)
+\nabla_\perp^2.
\ee

In this subsection, we will consider scalar fields and so-called 
semitransparent plates, 
that is, ones described by a $\delta$-function potential
\be
V=\lambda\delta(\xi-\xi_a), \ee
where $\xi_a$ is the position of the plate.
For a single semitransparent
plate at $\xi_1$, the Green's function can be written as
\be
G(x,x')=\int\frac{\rmd\omega}{2\pi}\frac{\rmd^2 k}{(2\pi)^2}\rme^{-\rmi
\omega(\tau-\tau')}
\rme^{\rmi\mathbf{k\cdot(r-r')_\perp}}g(\xi,\xi'),
\ee
where the reduced Green's function satisfies
\be
\left[-\frac{\omega^2}{\xi^2}-\frac1\xi\frac\partial{\partial\xi}\left(\xi\frac
\partial{\partial\xi}\right)+k^2+\lambda\delta(\xi-\xi_1)\right]g=
\frac1\xi\delta(\xi-\xi'),
\ee
which we recognize as just the
semitransparent cylinder problem with $m\to \zeta=-\rmi\omega$ and
$\kappa\to k$ \cite{CaveroPelaez:2006rt}.
The reduced Green's function for single plate thus is
\numparts
\bea
g(\xi,\xi')=I_\zeta(k\xi_<)K_\zeta(k\xi_>)-\frac{\lambda\xi_1 K_\zeta^2(k\xi_1)
I_\zeta(k\xi)I_\zeta(k\xi')}{1+\lambda\xi_1I_\zeta(k\xi_1)K_\zeta(k\xi_1)},
\qquad \xi,\xi'<\xi_1,\\
=I_\zeta(k\xi_<)K_\zeta(k\xi_>)-\frac{\lambda\xi_1 I_\zeta^2(k\xi_1)
K_\zeta(k\xi)K_\zeta(k\xi')}{1+\lambda\xi_1I_\zeta(k\xi_1)K_\zeta(k\xi_1)},
\qquad \xi,\xi'>\xi_1.
\eea 
\endnumparts
In the limit
$\lambda\to\infty$ we recover the well-known Green's function
subject to Dirichlet boundary conditions.

If we use the uniform asymptotic expansion (UAE), based on the limit
\bea
\xi\to\infty,&&\quad \xi_1\to\infty, \quad \xi-\xi_1 \mbox{ finite },
\quad\zeta=\hat\zeta\xi_1\to\infty,\quad \hat\zeta \mbox{ finite },
\eea
we recover the Green's function for a single plate in Minkowski space,
\be
\xi_1 g(\xi,\xi')\to  \frac{\rme^{-\kappa|\xi-\xi'|}}{2\kappa}
-\frac\lambda{2\kappa}\frac{\rme^{-\kappa(|\xi-\xi_1|+
|\xi'-\xi_1|)}}{\lambda+2\kappa},
\ee
where $\kappa=\sqrt{k^2+\zeta^2}$, $\omega=\rmi\zeta$.

The canonical energy-momentum for a scalar field is given by
\be
T_{\mu\nu}=\partial_\mu\phi\partial_\nu\phi
+g_{\mu\nu}\frac1{\sqrt{-g}}\mathcal{L},
\ee
where the Lagrange density includes the $\delta$-function potential.  Using
the equations of motion, we find the energy density to be
\bea
T_{00}=\frac12\left(\frac{\partial\phi}{\partial\tau}\right)^2-\frac12\phi
\frac{\partial^2}{\partial\tau^2}\phi+\frac\xi 2\frac\partial{\partial\xi}
\left(\phi\xi\frac\partial{\partial\xi}\phi\right)+\frac{\xi^2}2\bnabla_\perp
\cdot(\phi\bnabla_\perp\phi).\label{t00}\eea
From this and similar expressions for the spatial components of $T^{\mu\nu}$
we can obtain
$\langle T_{\mu\nu}\rangle$ by applying a differential operator to the
Green's function, in view of the one-loop connection \eref{vev}.

The force density is given by
\be
f_\lambda=-\frac1{\sqrt{-g}}\partial_\nu(\sqrt{-g}T^\nu{}_\lambda)
+\frac12T^{\mu\nu}\partial_\lambda g_{\mu\nu},
\ee
which follows from \eref{intbyparts},
or
\be
f_\xi=-\frac1\xi\partial_\xi(\xi T^{\xi\xi})-\xi T^{00}.\label{fxi}
\ee
When we integrate over all space to get the force, 
\be
F=\int(\rmd \mathbf{x})\sqrt{-g}f_\xi,
\ee
the first term in \eref{fxi} is
a surface term which does not contribute, as noted in \eref{intbyparts},
so the gravitational force on the system is
\be
\mathcal{F}=\int \rmd\xi \xi f_\xi=-\int\frac{\rmd\xi}{\xi^2}T_{00},\ee
which when multiplied by the gravitational acceleration $g$ is the
gravitational force/area on the Casimir energy.
Using the expression \eref{t00} for the energy density, and rescaling
we see that
\be
\mathcal{F}
=\int \rmd\xi \xi\int\frac{\rmd\hat\zeta \, \rmd^2k}{(2\pi)^3}\hat\zeta^2
g(\xi,\xi)\quad(\zeta=\hat\zeta\xi).\label{genforce}
\ee

This result is an immediate consequence of the general formula
\cite{Milton:2004ya}
\be
E_c=\frac1{2\rmi}\int(\rmd\mathbf{r})\int\frac{\rmd\omega}{2\pi}
2\omega^2\mathcal{G}(\mathbf{r,r}),
\ee
in terms of the frequency transform of the Green's function,
\be
G(x,x')=\int_{-\infty}^\infty\frac{\rmd\omega}{2\pi}\rme^{-i\omega(t-t')}
\mathcal{G}(\mathbf{r,r'}).
\ee

Alternatively, we can start from the following formula for the force
density for a single semitransparent plate,
\be
f_\xi=\frac12\phi^2\partial_\xi \lambda\delta(\xi-\xi_1),
\ee
or, in terms of the Green's function,
\be
\mathcal{F}=-\lambda\frac12
\int\frac{\rmd\zeta\,\rmd^2k}{(2\pi)^3}\partial_{\xi_1}[\xi_1 g(\xi_1,\xi_1)].
\ee

For example, the force on a single plate is given by
\be
\mathcal{F}=-\partial_{\xi_1} \frac12\int\frac{\rmd\zeta\,\rmd^2k}{(2\pi)^3}\ln
[1+\lambda I_\zeta(k\xi_1)K_\zeta(k\xi_1)],
\ee
Expanding this about some arbitrary point $\xi_0$,
with $\zeta=\hat\zeta\xi_0$,
and using the UAE, we get
\be
\mathcal{F}=-\frac1{96\pi^2 a^3}\int_0^\infty\frac{\rmd y\,y^2}{1+y/\lambda a},
\ee
which is just the negative of the 
(divergent) quantum vacuum energy of a single plate.

For two plates at $\xi_1$, $\xi_2$, for $\xi, \xi'<\xi_1<\xi_2$,
\numparts
\be
g(\xi,\xi')=I_<K_>
-\frac{\lambda_1\xi_1K_1^2+\lambda_2\xi_2K_1^2-\lambda_1\lambda_2\xi_1\xi_2
K_1K_2(K_2I_1-K_1I_2)}{\Delta}II_\prime,\ee
while for $\xi, \xi'>\xi_2>\xi_1$,
\be
g(\xi,\xi')=I_<K_>-\frac{\lambda_1\xi_1I_1^2+\lambda_2\xi_2I_2^2
+\lambda_1\lambda_2\xi_1\xi_2
I_1I_2(I_2K_1-I_1K_2)}{\Delta}KK_\prime,\ee
and between the plates, $\xi_1<\xi, \xi'<\xi_2$,
\bea
g(\xi,\xi')=I_<K_>
-\frac{\lambda_2\xi_2K_2^2(1+\lambda_1\xi_1K_1I_1)}\Delta II_\prime
-\frac{\lambda_1\xi_1I_1^2(1+\lambda_2\xi_2K_1I_1)}\Delta KK_\prime\nn\\
\mbox{}+\frac{\lambda_1\lambda_2\xi_1\xi_2I_1^2K_2^2}
\Delta(IK_\prime+KI_\prime),
\eea\endnumparts
where
\bea
\Delta={1+\lambda_1\xi_1K_1I_1+\lambda_2\xi_2I_2K_2-\lambda_1\lambda_2
\xi_1\xi_2I_1K_2(I_1K_2-I_2K_1)},
\eea
and where we have used the abbreviations
$I_a=I_\zeta(k\xi_a)$, $I=I_\zeta(k\xi)$, $I_\prime=I_\zeta(k\xi')$, etc.

In the weak acceleration limit, the Green's function
reduces to exactly the expected result, for $\xi_1<\xi,\xi'<\xi_2$
($a=\xi_2-\xi_1$)
\bea
\xi_0g(\xi,\xi')\to\frac1{2\kappa}\rme^{-\kappa|\xi-\xi'|}
+\frac1{2\kappa\tilde\Delta}\bigg[\frac{\lambda_1\lambda_2}
{4\kappa^2}2\cosh\kappa
(\xi-\xi')\nn\\
\mbox{}-\frac{\lambda_1}{2\kappa}\left(1+\frac{\lambda_2}{2\kappa}\right)
\rme^{-\kappa(\xi+\xi'-2\xi_2)}
-\frac{\lambda_2}{2\kappa}\left(1+\frac{\lambda_1}{2\kappa}\right)
\rme^{\kappa(\xi+\xi'-2\xi_1)}\bigg],
\eea
with
\be
\tilde\Delta=\left(1+\frac{\lambda_1}{2\kappa}\right)
\left(1+\frac{\lambda_2}{2\kappa}\right)\rme^{2\kappa a}-
\frac{\lambda_1\lambda_2}{4\kappa^2},\label{tdelta}
\ee
 The flat space limit also holds outside the plates.

In general, we have two alternative forms for the force on the two-plate
system:
\be
\mathcal{F}=-(\partial_{\xi_1}+\partial_{\xi_2})\frac12\int\frac{\rmd\zeta\,
\rmd^2k}{(2\pi)^3}\ln\Delta,\ee
which is equivalent to \eref{genforce}.
From either of these two formul\ae, we find the
gravitational force on the Casimir energy to be in the $\xi\to\infty$ limit
\be
\mathcal{F}=-\frac1{4\pi^2}\int_0^\infty \rmd\kappa\,\kappa^2 \ln\Delta_0
\ee
where $\Delta_0=\rme^{-2\kappa a}\tilde\Delta$, $\tilde\Delta$ given
in \eref{tdelta}.
From this we get the explicit force
\bea
\mathcal{F}&=&\frac1{96\pi^2 a^3}\int_0^\infty\rmd y\,y^3
\frac{1+\frac1{y+\lambda_1a}
+\frac1{y+\lambda_2a}}{\left(\frac{y}{\lambda_1a}+1\right)
\left(\frac{y}{\lambda_2a}+1\right)\rme^y-1}\nn\\
&&\mbox{}-\frac1{96\pi^2 a^3}\int_0^\infty \rmd y\,y^2
\left[\frac1{\frac{y}{\lambda_1a}
+1}+\frac1{\frac{y}{\lambda_2a}+1}\right]
=-(\mathcal{E}_c+\mathcal{E}_{d1}+\mathcal{E}_{d2}),
\eea
which is just the negative of the Casimir energy of the two semitransparent
plates, {\em including divergent parts associated with each
plate.}
The divergent terms simply renormalize the masses $m_i$ per area of each plate:
\bea
\mathcal{E}_{\rm total}=m_1+m_2+\mathcal{E}_{d1}+\mathcal{E}_{d2}+\mathcal{E}_c
=M_1+M_2+\mathcal{E}_c,
\eea
$M_i=m_i+\mathcal{E}_{di}$,
and thus the gravitational force on the entire apparatus obeys the
equivalence principle
\be
g\mathcal{F}=-g(M_1+M_2+\mathcal{E}_c).
\ee

\subsection{Centripetal Acceleration}
\begin{figure}
\centering
\vspace{2in}
\includegraphics{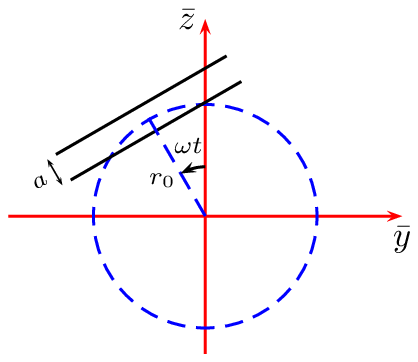}
 \includegraphics{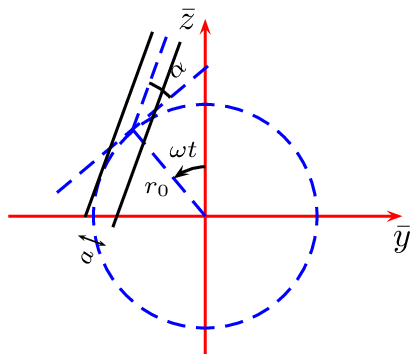}
\caption{\label{figcent1} Casimir apparatus undergoing uniform circular
motion with angular velocity $\omega$. The figure on the left
has the Casimir plates tangent to the circle of revolution, while
on the right the plates make an angle $\alpha$ with respect to that
tangent.}
\end{figure}

Consider finally a Casimir apparatus undergoing centripetal acceleration 
with angular velocity $\omega$
($\omega r\ll1$), as illustrated in \fref{figcent1}.
A detailed calculation \cite{Shajesh:2007sc} shows that the force
on the Casimir apparatus is just as expected, 
including the divergent pieces which renormalize the masses per area $A$
of the plates:
\bea
\frac{\mathbf{F}}A=-\frac{\omega^2}A
\int \rmd^3x\,\mathbf{r}\,t^{00}(\mathbf{r})
=-\omega^2\mathbf{r}_{\rm CM}(m_1+m_2+\mathcal{E}_{d1}+
\mathcal{E}_{d2}+\mathcal{E}_c),\eea
where $\mathbf{r}_{\rm CM}$ is the center of energy of the entire
system.
Nontrivial is the fact that this result is independent of the
orientation of the plates.

\section{Multiple Scattering Technique}
\label{sec3}
In the past few years, there has been a tremendous
resurgence of interest in the multiple-scattering technique.
This is largely inspired by the desire to improve the
comparison with experiment, such comparison heretofore having been restricted
to the use of the ``proximity force approximation,'' which
describes the Casimir interaction between bodies with curved
surfaces by pairwise interactions between small plane elements
of the surfaces \cite{blocki}.
The multiple-scattering 
method certainly dates back to at least the Krein formula
\cite{krein}, and was first used in connection with the Casimir
effect by Renne \cite{renne}.  A most famous variation were the
multiple reflection calculations of Balian and Duplantier \cite{balian}.
In the last few years there has
been an explosion of publications on variations of this
technique, as physicists realized it could be practically used
to obtain numerical results for forces between bodies of
arbitrary shapes, for example \cite{buscher,reynaud,bulgac,dalvit,
bordagx,kenneth1,kenneth,wirzba,bordag,maianeto,emig, Emig:2007me,
Emig:2007qw}.

We have also contributed to this redevelopment, giving the
general formulation for scalar fields \cite{Milton:2007wz},
and applying it to bodies described by $\delta$-function potentials
\cite{Milton:2007gy},
and, after generalizing to electrodynamics, to dielectric bodies
\cite{Milton:2008vr}.  The methodology has also been applied
to non-contact gears \cite{CaveroPelaez:2008tj,CaveroPelaez:2008tk}.

The multiple scattering approach starts from the well-known formula
for the vacuum energy or
Casimir energy (for simplicity here we first restrict attention to a
massless scalar field)($\tau$ is the ``infinite'' time that the
configuration exists) 
\be
E=\frac{\rmi}{2\tau}\mbox{Tr}\ln G\to\frac{\rmi}{2\tau}\mbox{Tr}\ln G G_0^{-1},
\ee
where $G$ ($G_0$) is the Green's function,
\be
(-\partial^2+V)G=1, \quad -\partial^2 G_0=1, \ee
subject to appropriate boundary conditions at infinity.  We will use
the Feynman or causal Green's functions.

Now we define the $T$-matrix \cite{lippmann}
\be
T=S-1=V(1+G_0V)^{-1}.\ee
If the potential has two disjoint parts, $V=V_1+V_2$
it is easy to derive
\be
T=(V_1+V_2)(1-G_0T_1)(1-G_0T_1G_0T_2)^{-1}(1-G_0T_2),\ee
where
\be
T_i=V_i(1+G_0V_i)^{-1},\quad i=1,2.\ee
Thus, we can write the general expression for 
the interaction between
the two bodies (potentials) in two equivalent forms:
\numparts
\bea
E_{12}&=& -\frac{\rmi}{2\tau}\mbox{Tr}\ln(1-G_0T_1G_0T_2)\\
&=&-\frac{\rmi}{2\tau}\mbox{Tr}\ln(1-V_1G_1V_2G_2),\eea
\endnumparts
where $G_i=(1+G_0V_i)^{-1}G_0,\quad i=1,2.$

\subsection{Multipole expansion}
\label{sec4.1}
To proceed to apply this method to general bodies, we use
an even older technique, the multipole expansion.  Let's illustrate
this with a $2+1$ dimensional version, which allows us to describe
cylinders with parallel axes.  We seek an expansion of the free Green's
function
\bea
G_0(\mathbf{R+r'-r})=\frac{\rme^{\rmi|\omega||\mathbf{r-R-r'}|}}{4\pi|
\mathbf{r-R-r'}|}
=\int\frac{\rmd k_z}{2\pi}
\rme^{\rmi k_z(z-Z-z')}g_0(\mathbf{r_\perp-R_\perp-r'_\perp}).
\eea
Here, the reduced Green's function is ($\omega=\rmi \zeta$)
\be
g_0(\mathbf{r_\perp-R_\perp-r'_\perp})=\int\frac{(\rmd^2k_\perp)}{(2\pi)^2}
\frac{\rme^{-\rmi\mathbf{k_\perp\cdot R_\perp}}\rme^{\rmi
\mathbf{k_\perp\cdot(r_\perp
-r'_\perp)}}}{k_\perp^2+k_z^2+\zeta^2}.\ee
As long as the two potentials do not overlap, so that we have
$\mathbf{r_\perp-R_\perp-r'_\perp}\ne0$, we can write an expansion in
terms of modified Bessel functions:
\bea
g_0(\mathbf{r_\perp-R_\perp-r'_\perp})=\sum_{m,m'}I_m(\kappa r) 
\rme^{\rmi m\phi}I_m'(\kappa r')\rme^{-\rmi m'\phi'}
\tilde g^0_{m,m'}(\kappa R),\quad\kappa^2=k_z^2+\zeta^2.\eea

By Fourier transforming, and using the definition of the Bessel function
\be
\rmi^mJ_m(kr)=\int_0^{2\pi}\frac{\rmd\phi}{2\pi}\,\rme^{-\rmi m\phi}
\rme^{\rmi kr\cos\phi},\ee
we easily find
\bea
\tilde g^0_{m,m'}(\kappa R)
=\frac1{2\pi}\int \frac{\rmd k\,k}{k^2+\kappa^2}J_{m-m'}(kR)
\frac{J_m(kr)J_m(kr')}{I_m(\kappa r)I_m(\kappa r')}
=\frac{(-1)^{m'}}{2\pi}K_{m-m'}(\kappa R).\eea
Thus we can derive an expression for the interaction between two bodies, in
terms of discrete matrices,
\be \mathfrak{E}\equiv
\frac{E_{\rm int}}{L}=\frac1{8\pi^2}\int\rmd\zeta\,\rmd k_z\ln\det
\left(1-\tilde g^0
\tilde T_1\tilde g^{0\top} \tilde T_2\right),\ee
where $\top$ denotes transpose, and the $\tilde T$ matrix elements are given by
\bea
\tilde T_{mm'}=\int \rmd r\,r\,\rmd\phi \int \rmd r'\,r'\,\rmd\phi'
I_m(\kappa r)
\rme^{-\rmi m\phi}I_{m'}(\kappa r')\rme^{\rmi m'\phi'}T(r,\phi;r',\phi').\eea

\begin{figure}
\centering
\vspace{2.0in}
\includegraphics{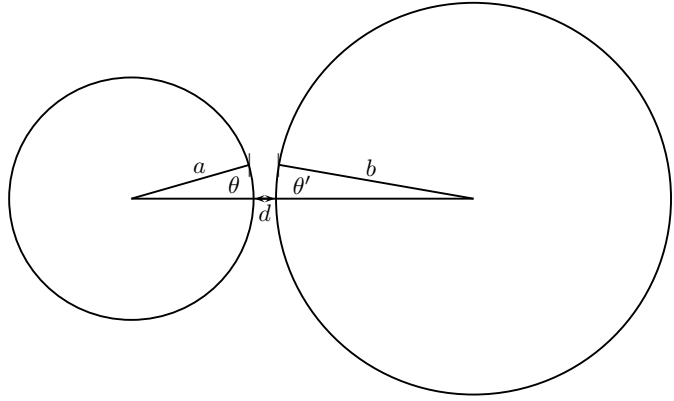}
\caption{Geometry of two cylinders (or two spheres) with radii $a$ and
$b$, respectively, and distances between their centers of $R>a+b$.
The proximity force approximation applies when the distance of closest
approach $d=R-a-b\ll a,b$.  The approximation consists in assuming
that the interaction is dominated by the interaction of adjacent surface
elements, as shown.}
\label{pfa}
\end{figure}

We will apply this 2+1 dimensional formalism to the interaction between
two cylinders, as illustrated in \fref{pfa}.  Also illustrated there is
the idea behind the proximity force approximation, which applies when
the separation between the bodies is much smaller than the radii of
curvature of the bodies.

Consider two parallel semitransparent cylinders, of
radii $a$ and $b$, respectively, lying
outside each other, described by the potentials
\be
V_1=\lambda_1 \delta(r-a),\quad V_2=\lambda_2\delta(r'-b),
\ee
with the separation between the centers $R$ satisfying $R>a+b$.
It is easy to work out the scattering matrix in this situation,
\be
(t_1)_{mm'}=2\pi\lambda_1a\delta_{mm'}\frac{I_m^2(\kappa a)}{1+\lambda_1 a
I_m(\kappa a)K_m(\kappa a)}.\label{met}
\ee
Thus the Casimir energy per unit length is
\be
\mathfrak{E}=\frac1{4\pi}\int_0^\infty \rmd\kappa\,\kappa\,\mbox{tr}\ln(1-A),
\label{eylenergy}
\ee
where $A=B(a)B(b)$,
in terms of the matrices
\be
B_{mm'}(a)=K_{m+m'}(\kappa R)\frac{\lambda_1 a I_{m'}^2(\kappa a)}
{1+\lambda_1 aI_{m'}(\kappa a)K_{m'}(\kappa a)}.\label{bofa}
\ee

The strong coupling limit, the case of Dirichlet cylinders, has been
considered in \cite{Emig:2007me}, for example.  In contrast, here we 
will consider the situation of
weak coupling, $\lambda_1a\ll 1$, $\lambda_2b\ll1$, where the formula
for the interaction energy between two cylinders becomes
\bea
\mathfrak{E}=
-\frac{\lambda_1\lambda_2 ab}{4\pi R^2}\sum_{m,m'=-\infty}^\infty
\int_0^\infty \rmd x\,x \,K_{m+m'}^2(x) 
I_m^2(xa/R)I_{m'}^2(xb/R).\label{wcenergy}
\eea
One merely exploits the small argument
expansion for the modified Bessel functions $I_m(xa/R)$ and $I_{m'}(xb/R)$:
\be
I_m^2(x)=\left(\frac{x}{2}\right)^{2|m|}
\sum_{n=0}^\infty Z_{|m|,n} \left(\frac{x}{2}\right)^{2n},
\ee
where the coefficients $Z_{m,n}$ are
\be
Z_{m,n} = \frac{2^{2(m+n)} \;
\Gamma  \left( m+n+\frac{1}{2} \right)}{\sqrt{\pi}\;
n! \; (2m+n)! \; \Gamma( m+n+1 ) }.
\ee
In this case we get an amazingly simple result
\be
\mathfrak{E}=-\frac{\lambda_1a\lambda_2b}{4\pi R^2}\frac12\sum_{n=0}^\infty
\left(\frac{a}{R}\right)^{2n}P_n(\mu),\label{multicyl}
\ee
where $\mu=b/a$, and
where by inspection we identify the binomial coefficients
\be
P_n(\mu)=\sum_{k=0}^n {n\choose{k}}^2
\mu^{2k}.
\ee
Remarkably, it is possible to perform the sums,
so we obtain the following
closed form for the interaction between two weakly-coupled cylinders:
\be
\mathfrak{E}=-\frac{\lambda_1 a \lambda_2 b}{8 \pi R^2}
\left[\!\!\left( 1 - \left( \frac{a+b}{R} \right)^2\right)\!\!
\left( 1 - \left( \frac{a-b}{R} \right)^2 \right)\!\!\right]^{-1/2}.
\label{excyl}
\ee

We note that in the limit $R-a-b=d\to0$, $d$ being the distance between
the closest points on the two cylinders, we recover the proximity force
theorem in this case 
\be
 U(d)=-\frac{\lambda_1\lambda_2 }{32\pi}\sqrt{\frac{2ab}R}\frac1{d^{1/2}},
\qquad d\ll a, b.
\label{pfawccyl1}
\ee
The rate of approach is 
given by
\be
\frac{\mathfrak{E}}U\approx1-\frac{1+\mu+\mu^2}{4\mu}\frac{d}R
\approx 1-\frac{R^2-aR+a^2}{4a(R-a)}\frac{d}R.\label{ratio}
\ee

\begin{figure}[t]
\centering
\vspace{2.in}
\includegraphics{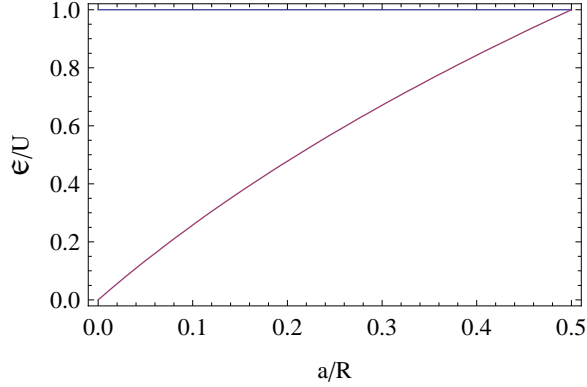}
\caption{\label{figwccyl1}
Plotted is the ratio of the exact interaction energy
 of two equal radii weakly-coupled
cylinders to the proximity force approximation.}
\end{figure}

\begin{figure}[t]
\centering
\vspace{2in}
\includegraphics{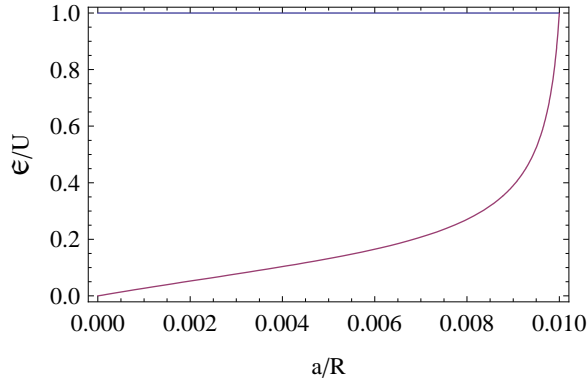}
\caption{\label{figwccyl2}
Plotted is the ratio of the exact interaction energy
of two weakly-coupled
cylinders with ratio of radii $b/a=99$ to the proximity force approximation.}
\end{figure}

The comparison between the exact result, and the proximity force
approximation \eref{pfawccyl1} is given in \fref{figwccyl1},
for equal radii, and in \fref{figwccyl2} for unequal radii.
Evidently, the proximity force approximation fails badly as the
cylinders are pulled apart.

\begin{figure}[t]
\vspace{2.in}
\includegraphics{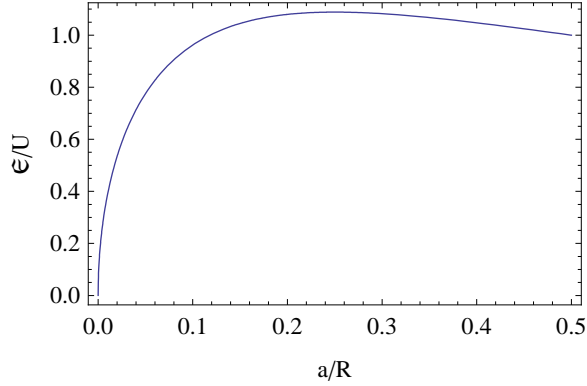}
\caption{Interaction energy between a weakly coupled $\delta$-function
cylinder and a Dirichlet plane.  Plotted is the ratio between the energy
and the PFA approximation \eref{pfa:cylpl}\label{figcylpl}.}
\end{figure}

We can similarly consider the interaction
between a cylinder and a plane.  By the method
of images, we can find the
interaction between a semitransparent cylinder and a Dirichlet
plane to be
\be
\mathfrak{E}=\frac1{4\pi}\int_0^\infty \kappa\,\rmd\kappa\,\mbox{tr}\,
\ln(1-B(a)),\label{cylpl}\ee
where $B(a)$ is given in \eref{bofa}.
In the strong-coupling limit this result agrees with that given by Bordag
and Nikolaev \cite{bordag}, because
\be\mbox{tr}\,B^s=\mbox{tr}\,\tilde B^s,
\quad \tilde B_{mm'}=\frac1{K_m(\kappa a)}K_{m+m'}(
\kappa R) I_{m'}(\kappa a).
\ee
In exactly the way given above, we can obtain a
closed-form result for the interaction energy between a Dirichlet plane
and a weakly-coupled cylinder of radius $a$
separated by a distance $R/2$.  The result is again quite
simple:
\be
\mathfrak{E}=-\frac{\lambda a}{4\pi R^2}\left[1-
\left(\frac{2a}R\right)^2\right]^{-3/2}.\label{excylpl}
\ee
In the limit as $d\to0$, this agrees with the PFA for this situation:
\be
U(d)=-\frac{\lambda}{64\pi}\frac{\sqrt{2a}}{d^{3/2}}.\label{pfa:cylpl}
\ee
Note again that this form is ambiguous: the proximity force theorem is
equally well satisfied if we replace $a$ by $R/2$, for example, in $U(d)$.
The exact result is compared with the PFA in \fref{figcylpl}.

\subsection{3-dimensional formalism}
\label{sec4.2}
The three-dimensional formalism is very similar.  In this case, the
free Green's function has the representation
\bea
G_0(\mathbf{R+r'-r})&=&\sum_{lm,l'm'}j_l(\rmi|\zeta|r)j_{l'}(\rmi|\zeta|r')
Y_{lm}^*(\mathbf{\hat r})
Y_{l'm'}(\mathbf{\hat r'})g_{lm,l'm'}(\mathbf{R}).\eea
The reduced Green's function can be written in the form
\bea
g^0_{lm,l'm'}(\mathbf{R})
=(4\pi)^2\rmi^{l'-l}\int\frac{(\rmd\mathbf{k})}
{(2\pi)^3}\frac{\rme^{\rmi\mathbf{k\cdot R}}}{k^2+\zeta^2}
\frac{j_l(kr)j_{l'}(kr')}
{j_l(\rmi|\zeta| r)j_{l'}(\rmi|\zeta|r')}
Y_{lm}(\mathbf{\hat k})Y^*_{l'm'}(\mathbf{\hat k}).
\eea

Now we use the plane-wave expansion 
\be
\rme^{\rmi\mathbf{k\cdot R}}=4\pi\sum_{l''m''}\rmi^{l''}j_{l''}(kR)
Y_{l''m''}(\mathbf{\hat R})Y_{l''m''}^*(\mathbf{\hat k}),
\ee
so now we encounter something new, an integral over three spherical
harmonics,
\be
\int \rmd\mathbf{\hat k}Y_{lm}(\mathbf{\hat k})Y_{l'm'}^*(\mathbf{\hat k})
Y_{l''m''}^*(\mathbf{\hat k})=C_{lm,l'm',l''m''},\ee
where
\bea
C_{lm,l'm',l''m''}=(-1)^{m'+m''}\sqrt{\frac{(2l+1)(2l'+1)(2l''+1)}{4\pi}}
\left(\begin{array}{ccc}
l&l'&l''\\
0&0&0\end{array}\right)
\left(\begin{array}{ccc}
l&l'&l''\\
m&m'&m''\end{array}\right).
\eea
The three-$j$ symbols (Wigner coefficients)
here vanish unless $l+l'+l''$ is even. 
 This fact
is crucial, since because of it we can
write $j_{l''}(kR)$ in terms of Hankel functions of the first and second
kind, using the reflection property of the latter,
\be
h_{l''}^{(2)}(kR)=(-1)^{l''}h^{(1)}_{l''}(-kR),\ee
and then extending the $k$ integral over the entire real axis to a
contour integral closed in the upper half plane.  This leads to
\bea
g^0_{lm,l'm'}(\mathbf{R})=4\pi  \rmi^{l'-l}\sqrt{\frac{2|\zeta|}{\pi R}}
\sum_{l''m''}C_{lm,l'm',l''m''}K_{l''+1/2}(|\zeta|R)
Y_{l''m''}(\mathbf{\hat R}).
\eea

For the case of two semitransparent spheres that are totally outside
each other,
\be
V_1(r)=\lambda_1\delta(r-a),\qquad V_2(r')=\lambda_2\delta(r'-b),
\ee
in terms of spherical coordinates centered on each sphere, it is again
very easy to calculate the scattering matrices,
\bea
T_1(\mathbf{r,r'})=\frac{\lambda_1}{a^2}\delta(r-a)\delta(r'-a)
\sum_{lm}\frac{Y_{lm}(\mathbf{\hat r})Y_{lm}^*(\mathbf{\hat r'})}{1+\lambda_1a
K_{l+1/2}(|\zeta|a)I_{l+1/2}(|\zeta|a)},\eea
and then the harmonic transform is very similar to that seen for
the cylinder, ($k=\rmi|\zeta|$)
\bea
(t_1)_{lm,l'm'}&=&\int(\rmd\mathbf{r})(\rmd\mathbf{r'})
j_l(kr)Y^*_{lm}(\mathbf{\hat r})
j_{l'}(kr')Y_{l'm'}(\mathbf{\hat r'})T_1(\mathbf{r,r'})\nn\\
&=&\delta_{ll'}\delta_{mm'}(-1)^l\frac{\lambda_1 a\pi}{2|\zeta|}
\frac{I_{l+1/2}^2(|\zeta|a)}{1+\lambda_1 a K_{l+1/2}(|\zeta|a)
I_{l+1/2}(|\zeta|a)}.\eea
Let us suppose that the two spheres lie along the $z$-axis, that is,
$\mathbf{R}=
R\mathbf{\hat z}$.  Then we can simplify the expression for the energy
somewhat by using $Y_{lm}(\theta=0)=\delta_{m0}\sqrt{(2l+1)/4\pi}$.
The formula for the energy of interaction becomes
\be
E=\frac1{2\pi}\int_0^\infty \rmd\zeta \,\mbox{tr}\ln(1-A),\label{etrace}
\ee
in terms of the matrix
\be
A_{lm,l'm'}=\delta_{m,m'}\sum_{l''}B_{ll''m}(a)B_{l''l'm}(b)
\ee
with
\bea
B_{ll'm}(a)&=&
\frac{\sqrt{\pi}}{\sqrt{2\zeta R}}\rmi^{-l+l'} \sqrt{(2l+1)(2l'+1)}
\sum_{l''}(2l''+1)\nn\\
&&\times\left(\begin{array}{ccc}l&l'&l''\\0&0&0\end{array}
\right)
\left(\begin{array}{ccc}l&l'&l''\\m&-m&0\end{array}\right)
\frac{K_{l''+1/2}(\zeta R)\lambda_1 a I_{l'+1/2}^2(\zeta a)}
{1+\lambda_1 a I_{l'+1/2}(\zeta a)K_{l'+1/2}(\zeta a)}.
\eea
Note that the phase always cancels in the trace.

Again, the strong coupling limit was considered in \cite{Emig:2007me}.
For weak coupling, a major simplification results because of the
orthogonality property,
\be
\sum_{m=-l}^l\left(\begin{array}{ccc}l&l'&l''\\m&-m&0\end{array}
\right)\left(\begin{array}{ccc}l&l'&l'''\\m&-m&0\end{array}\right)
=\delta_{l''l'''}\frac1{2l''+1},\qquad l\le l',
\ee and then we find for the interaction energy
\bea
E=-\frac{\lambda_1a\lambda_2b}{4R}\int_0^\infty \frac{dx}x\sum_{ll'l''}
(2l+1)(2l'+1)(2l''+1)
\left(\begin{array}{ccc}l&l'&l''\\0&0&0\end{array}
\right)^2\nn\\
\times K_{l''+1/2}^2(x)I_{l+1/2}^2(xa/R)I_{l'+1/2}^2(xb/R).
\eea

As with the cylinders, we expand the modified Bessel functions
of the first kind in power series in $a/R,b/R<1$. This expansion yields
the infinite series
\be
E=- \frac{\lambda_1 a \lambda_2 b }{4 \pi R}
\frac{a b}{R^2}
\sum_{n=0}^\infty \frac{1}{n+1} \sum_{m=0}^n
D_{n,m}\left(\frac{a}{R}\right)^{2(n-m)}\!\!\left(\frac{b}{R}\right)^{2m},
\label{mesphere}
\ee
where by inspection of the first several $D_{n,m}$ coefficients
we can identify them as
\be
D_{n,m}=\frac{1}{2} {2n+2\choose{2m+1}},
\ee
and now we can immediately sum the expression
 for the Casimir interaction energy to give the closed form
\be
E=\frac{\lambda_1 a \lambda_2 b}{16 \pi R} \; \ln\left(
\frac{1-\left(\frac{a+b}{R}\right)^2}{1-
\left(\frac{a-b}{R}\right)^2}\right).\label{exsphere}
\ee

\begin{figure}[t]
\centering
\vspace{2in}
\includegraphics{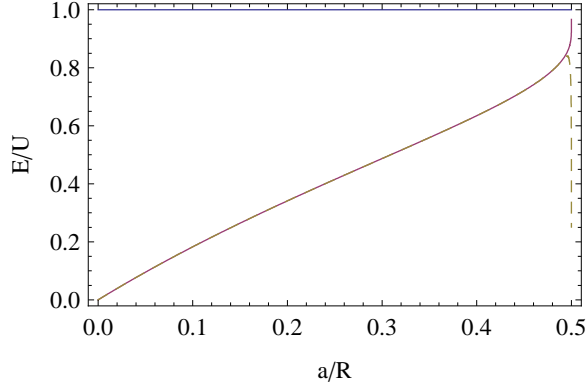}
\caption{\label{figspsp}
Plotted is the ratio of the exact interaction energy
of two weakly-coupled
spheres of equal radii to the proximity force approximation.}

\end{figure}

\begin{figure}
\centering
\vspace{2in}
\includegraphics{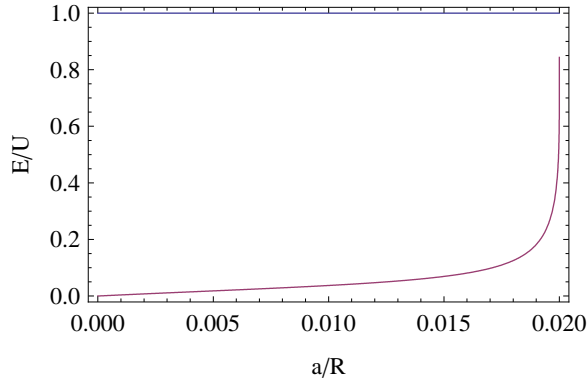}
\caption{\label{figspsp2}
Plotted is the ratio of the exact interaction energy
 of two weakly-coupled
spheres, with ratio of radii $b/a=49$, to the proximity force approximation.}

\end{figure}

Again, when $d=R-a-b\ll a,b$, the proximity force theorem 
is reproduced:
\be
U(d)\sim \frac{\lambda_1\lambda_2ab}{16\pi R}\ln (d/R),\qquad
d\ll a, b.\label{pfawcsphere1}
\ee
However, as \fref{figspsp} and \fref{figspsp2} demonstrate, the
approach is not very smooth,
even for equal-sized spheres. The ratio of the energy to the PFA is
($b/a=\mu$)
\be
\frac{E}{U}=1+\frac{\ln[(1+\mu)^2/2\mu]}{\ln d/R},\qquad d\ll a, b.
\ee

\begin{figure}[t]

\vspace{2in}
\centering
\includegraphics{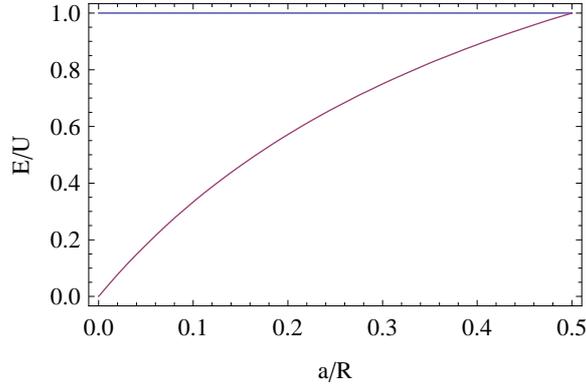}
\caption{\label{fig:exsppl}
Plotted is the ratio of the exact interaction energy
 of a weakly-coupled
sphere above a Dirichlet plane to the PFA.}

\end{figure}

In just the way indicated above, we can obtain a closed-form result
for the interaction energy between a weakly-coupled sphere and a Dirichlet
plane.  Using the simplification that
\be
\sum_{m=-l}^{l} (-1)^m
\left(\begin{array}{ccc}l&l&l'\\m&-m&0\end{array}\right)
\left(\begin{array}{ccc}l&l&l'\\0&0&0\end{array}\right)
=\delta_{l'0},
\ee
we can write the interaction energy in the form
\be
E=-\frac{\lambda a}{2 \pi R} \int_{0}^{\infty}
\rmd x \sum_{l=0}^{\infty}
\sqrt{\frac{\pi}{2x}} (2l+1) K_{1/2}(x)
I_{l+1/2}^2\left(x\frac{a}{R}\right).
\ee
Then in terms of $R/2$ as the distance between the center of the sphere and the
plane, the exact interaction energy is
\be
E=-\frac\lambda{2\pi}\left(\frac{a}R\right)^2\frac1{1-(2a/R)^2},\label{exsppl}
\ee
which as $a\to R/2$ reproduces the proximity force limit, contained in the
(ambiguously defined) PFA formula
\be
U=-\frac\lambda{8\pi}\frac{a}{d}.\label{pfa:sppl}
\ee
The ratio of the exact interaction to the PFA is
plotted in \fref{fig:exsppl}.

\subsection{Exact Results---Weak Coupling}
In weak coupling it is possible to derive the exact (scalar) interaction
between two potentials in general forms,
\numparts
\bea
2D:&\quad\mathfrak{E}&=\frac{E}{L}=-\frac1{32\pi^3}\int(\rmd
\mathbf{r_\perp})
(\rmd\mathbf{r_\perp'})
\frac{V_1(\mathbf{r_\perp})V_2(\mathbf{r_\perp'})}{|\mathbf{r-r'}|^2},\\
3D:& \quad E&=-\frac1{64\pi^3}\int(\rmd\mathbf{r})(\rmd\mathbf{r'})
\frac{V_1(\mathbf{r})V_2(\mathbf{r'})}{|\mathbf{r-r'}|^3}.\eea
\endnumparts
The results found above in sections \ref{sec4.1} and \ref{sec4.2} 
can be directly derived from these pairwise summation formul\ae. 
For exact results for $\delta$-function plates, see the contribution
to this Proceedings by Jef Wagner.

It is straightforward to adapt the multiple scattering formalism
to the far more relevant situation of electromagnetism.  For weak
coupling, this reduces to 
the (retarded dispersion) van der Waals (vdW) potential between polarizable
molecule given by
\be
V=-\frac{23}{4\pi}\frac{\alpha_1\alpha_2}{r^7},\qquad \alpha=\frac{\varepsilon
-1}{4\pi N}.\ee
This allows us to consider in the same vein (electromagnetic)
interactions between distinct dilute dielectric bodies of arbitrary shape.
(Of course, this potential represents the retarded limit of the
van der Waals force.  The non-retarded limit of the van der Waals force
between bodies has long been studied, for example between dielectric
cylinders \cite{mahanty,nyland}, and between a sphere and a plane
\cite{noguez}.) 

This vdW potential may be directly derived from the multiple
scattering formula
\be
W=-\frac{\rmi}2 \mbox{Tr}\,\ln\frac{\bGamma}{\bGamma_0}\approx- \frac{\rmi}2
\mbox{Tr}\,V_1\bGamma_0 V_2\bGamma_0,\ee
where where have omitted self-action terms, and
\be
\bGamma_0
=\bnabla\times\bnabla\times\mathbf{1}\frac{e^{-|\zeta||\mathbf{r-r'}|}}
{4\pi|\mathbf{r-r'}|}+\mathbf{1}.\ee
The quantity $V=\varepsilon-1$ is the potential representing the dielectric
body. 

\begin{figure}[h]
\centering
\includegraphics{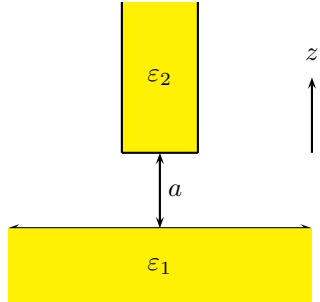}
\caption{\label{pl-pl} A plate of cross section $A$, a distance $a$ above
an infinite plate. Both plates are of semi-infinite extent in the direction
perpendicular to both plates.}
\end{figure}

We first apply this approach to compute the force between
a dilute dielectric slab and an infinite dilute dielectric  plate, as
shown in \fref{pl-pl}.
If the cross sectional area of the finite slab is $A$, the force between
the slabs is
\be
\frac{F}A=-\frac{23}{640\pi^2}\frac1{a^4}(\varepsilon_1-1)(\varepsilon_2-1),
\ee
which is the Lifshitz formula for infinite (dilute) slabs.
 Note that there is no correction due to the finite area of the
upper slab.

\begin{figure}[h]
\centering
\includegraphics{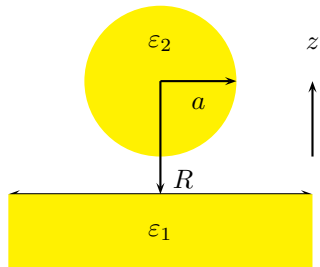}
\caption{\label{sp-pl} A dilute dielectric sphere, of radius $a$, 
the center of which is a
distance $Z$ above an infinite dilute dielectric plate.}
\end{figure}

In the same way we can immediately compute the force between a dilute
dielectric sphere of radius $a$ a distance $Z$ away from 
an infinite slab, as illustrated in \fref{sp-pl}.
The interaction energy of this system is
\be
E=-\frac{23}{640\pi^2}(\varepsilon_1-1)(\varepsilon_2-1)\frac{4\pi a^3/3}{Z^4}
\frac1{(1-a^2/Z^2)^2},\label{esphpl}\ee
which agrees with the PFA in the short separation limit, $Z-a=\delta\ll a$,
where the force is
\be
F_{\rm PFA}=2\pi a\mathcal{E}_\|(\delta)=-\frac{23}{640\pi^2}(\varepsilon_1-1)
(\varepsilon_2-1)\frac{2\pi a}{3\delta^3},\ee
with an exact correction, intermediate between that for scalar
averaged Dirichlet and Neumann \cite{wirzba,bordag} and electromagnetic 
perfectly-conducting boundaries \cite{maianeto}.

\begin{figure}
\centering
\includegraphics{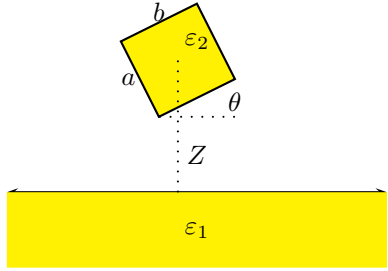}
\caption{\label{tiltedslab} Rectangular solid (with dielectric
constant $\varepsilon_2$) of side $a$, $b$, and $c$ (perpendicular
to the plane, not shown)
 a distance $Z$ above an infinite plate (with dielectric
constant $\varepsilon_1$) extending to $z=-\infty$.
The side $b$ makes an angle $\theta$ with respect to the plate.}
\end{figure}

\begin{figure}
\centering
\includegraphics{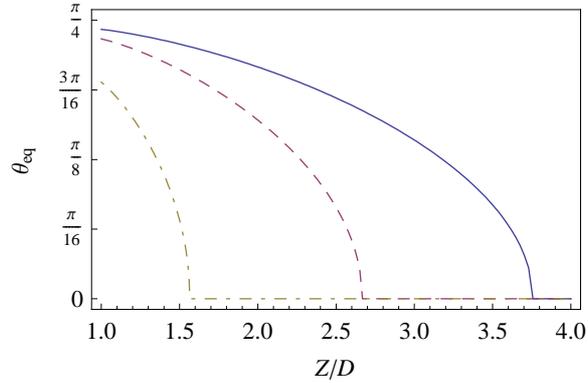}
\caption{\label{eqtilt} 
 The stable equilibrium
angle of a slab above an infinite plate for given $b/a$ ratios
0.95, 0.9, and 0.7, respectively given by solid, dashed, and dot-dashed
lines.}  
\end{figure}

We can also consider torques between bodies.  For example, consider
a rectangular dilute dielectric solid, with sides $a$, $b$, and $c$,
the center of mass of which is a distance $Z$ above a dilute dielectric
plate, but which is tilted through an angle $\theta$ relative
to the plate, as shown in \fref{tiltedslab}.
Generically, the shorter
side wants to align with the plate, which is obvious geometrically,
since that (for fixed center of mass position) minimizes the energy.
However, if the slab has square cross section, the equilibrium
position occurs when a corner is closest to the plate, also obvious
geometrically.   But if the two sides are close enough in length,
a nontrivial equilibrium position between these extremes can
occur, as well illustrated in \fref{eqtilt}.
For large enough separation, the shorter side wants to face the plate,
but for 
\be Z<Z_0=\frac{a}2\sqrt{\frac{2a^2+5 b^2
+\sqrt{9 a^4+20 a^2 b^2+20  b^4}}{5 \left(a^2-b^2\right)}},\qquad a>b,\ee
the equilibrium angle increases, until finally at
$Z=D=\sqrt{a^2+b^2}/2$ the slab touches the plate at an angle 
$\theta=\arctan b/a$, that is, the center of mass is just above the point of
contact, about which point there is no torque.

Finally, once again we can get exact results for the vdW interaction
 between parallel cylinders (radii $a,b$, center separation $R$):
\bea
\mathfrak{E}=
\frac{E}L=-\frac{23}{60\pi}(\varepsilon_1-1)(\varepsilon_2-1)\frac{a^2b^2}
{R^6}\frac{1-\frac12\left(\frac{a^2+b^2}{R^2}\right)-\frac12
\left(\frac{a^2-b^2}{R^2}\right)^2}{\left[\left(1-\left(\frac{a+b}R\right)^2
\right)\left(1-\left(\frac{a-b}R\right)^2\right)\right]^{5/2}},
\eea
and between spheres:
\bea
E=-\frac{23}{1920\pi}\frac{(\varepsilon_1-1)(\varepsilon_2-1)}{R}
\Bigg\{\ln\left(\frac{1-\left(\frac{a-b}R\right)^2}{1-
\left(\frac{a+b}R\right)^2}\right)\nn\\
\mbox{}+\frac{4ab}{R^2}\frac{\frac{a^6-a^4b^2-a^2b^4+b^6}{R^6}-
\frac{3a^4-14a^2b^2+3b^4}{R^4}+3\frac{a^2+b^2}{R^2}-1}{\left[\left(
1-\left(\frac{a-b}R\right)^2\right)\left(1-\left(\frac{a+b}R\right)^2\right)
\right]^2}\Bigg\}.
\eea
This last expression, which is rather ugly, may be verified to yield
the proximity force theorem:
\be
E\to U=-\frac{23}{640\pi}\frac{a(R-a)}{R\delta^2},\quad \delta=R-a-b\ll a,b.
\ee
It also, in the limit $b\to\infty$, $R\to\infty$ with $R-b=Z$ held
fixed, reduces to the result \eref{esphpl} for the interaction of a sphere
with an infinite plate.

The same methodology may be used to derive the force
between corrugated planes and corrugated concentric cylinders,
which calculations may have relevance to nanomechanics.  See
the contribution to this Proceedings by Cavero-Pel\'aez,
and \cite{CaveroPelaez:2008tj,CaveroPelaez:2008tk}.



\section{Comments and Prognosis}
In this talk I have described some recent exciting progress in
the practical development of quantum vacuum energy.  I have 
concentrated on the work of the Oklahoma group, without meaning
to denigrate the work of many others of equal or greater
importance. I will summarize the status of the topics I discussed
briefly here: 
\begin{itemize}
\item The temperature controversy can only be resolved 
by dedicated experiments,
designed to measure Casimir forces at different
temperatures  between metal surfaces 
\item The semiconductor anomaly needs to be examined more closely
both theoretically and experimentally.
\item Casimir energies, including their divergent contributions,
exhibit the inertial and gravitational masses expected:   $m=E/c^2$.
These gravitational interactions need to be examined for more
complicated geometries, such as for the Casimir energy of a 
perfectly conducting spherical shell, with the hope of understanding
the physical significance of the divergent contributions.
\item The multiple-scattering methods recently developed  are in fact not 
particularly novel, and illustrate
the ability of physicists to continually rediscover old methods.
What is new is the recognition that one can
evaluate continuum determinants (or infinitely
dimensional discrete ones) accurately numerically, and in some
cases even exactly in closed form.
This is making it possible to compute Casimir forces and torques for geometries
previously inaccessible.
\item It is indeed remarkable, if perhaps not surprising in retrospect,
to see that  closed form expressions can be obtained for
the interaction between spheres,  between parallel cylinders,
and for other geometries in weak
coupling. These results demonstrate most conclusively the unreliability of the
proximity force approximation (of course, the proximity force theorem 
holds true).
\item This methodology has been used to obtain new results for
non-contact gears: See the talk by In\'es Cavero-Pel\'aez.
\item Jef Wagner has presented new results for edge effects for semitransparent
potentials.
\item Undoubtedly these developments  will lead to improved conceptual 
understanding, and to
better comparison with experiment.
\end{itemize}

\ack
We thank the US National Science Foundation, grant number PHY-0554826,
and the US Department of Energy, grant number DE-FG-04ER41305, for
partial support of the research reported here.  We especially thank
Victor Dodonov for organizing such a successful meeting on ``60 Years
of the Casimir Effect,'' and the International Center for Condensed
Matter Physics at the University of Brasilia, 
and the College of Arts and Sciences and
the office of the Vice President for Research at
the University of Oklahoma for travel support.  
I thank my collaborators for allowing me to report on our joint work.

\end{document}